\documentclass[12pt]{article}
\usepackage{latexsym}
\def\hybrid{\topmargin -20pt    \oddsidemargin 0pt
        \headheight 0pt \headsep 0pt
       \textwidth 6.25in       
       \textheight 9.5in       
        \marginparwidth .875in
        \parskip 5pt plus 1pt   \jot = 1.5ex}

\hybrid
\makeatletter
\@addtoreset{equation}{section}
\makeatother

\newcommand{\color}[6]{}

\newcommand{\cL}{{\cal L}}

\newcommand{\cN}{{\cal N}}
\newcommand{\cO}{{\cal O}}
\newcommand{\cP}{{\cal P}}

\newcommand{\cW}{{\cal W}}

\newcommand{\hf}{\frac12}
\newcommand{\qt}{\frac14}
\newcommand{\bea}{\begin{eqnarray}}
\newcommand{\eea}{\end{eqnarray}}
\newcommand{\be}{\begin{equation}}
\newcommand{\ee}{\end{equation}}
\newcommand{\bt}{\begin{tabular}}
\newcommand{\et}{\end{tabular}}
\newcommand{\ba}{\begin{array}}
\newcommand{\ea}{\end{array}}

\def\IB{\relax{\rm I\kern-.18em B}}
\def\ID{\relax{\rm I\kern-.18em D}}
\def\IE{\relax{\rm I\kern-.18em E}}
\def\IF{\relax{\rm I\kern-.18em F}}
\def\IH{\relax{\rm I\kern-.18em H}}
\def\II{\relax{\rm I\kern-.18em I}}
\def\IK{\relax{\rm I\kern-.18em K}}
\def\IL{\relax{\rm I\kern-.18em L}}
\def\IM{\relax{\rm I\kern-.18em M}}
\def\IN{\relax{\rm I\kern-.18em N}}
\def\IP{\relax{\rm I\kern-.18em P}}
\def\IR{\relax{\rm I\kern-.18em R}}
\def\IT{\relax{\rm I\kern-.42em T}}
\def\IZ{\relax{\hbox{\raisebox{.38ex}
    {\scriptsize\bfseries\slshape /}\kern-.40em\_\kern-.28em\rm Z}}}
\def\Iz{\relax{\hbox{\raisebox{.38ex}
    {\tiny\bfseries\slshape /}\kern-.25em\raisebox{.65ex}
    {\tiny\bfseries\slshape /}\kern-.43em\_\kern-.26em\rm Z}}}
\def\inbar{\vrule height1.5ex width.8pt depth-0.2pt}
\def\inbarhi{\vrule height1.55ex width.5pt depth-.85ex}
\def\inbarlo{\vrule height.8ex width.5pt depth0ex}
\def\IC{\relax{\rm C\kern-.48em \inbar\kern.48em}}
\def\IO{\relax{\rm O\kern-.56em \inbar\kern.56em}}
\def\IQ{\relax{\rm Q\kern-.56em \inbar\kern.56em}}
\def\IS{\relax{\rm S\kern-.37em \inbarhi\kern.08em\inbarlo\kern.29em}}

\def \one{\relax{\rm 1\kern-.26em I}}

\def \soll={\stackrel{!}{=}}

\def \barfill{\leaders\hrule height 0.1 true pt\hfill}
\def \overbar#1{\vbox{\ialign{##\crcr\barfill\crcr\noalign{\kern 1pt
                                      \nointerlineskip}$\hfil{#1}\hfil$\crcr}}}
\def \scriptbar#1{{\vbox{\ialign{##\crcr\thinspace\barfill\thinspace\crcr
    \noalign{\kern 0.8pt\nointerlineskip}$\hfil{\scriptstyle #1}\hfil$\crcr}}}}

\newlength{\oldindent}
\newlength{\quadlength} 
\newlength{\abstand} \newlength{\breite}

\def \Nucl#1{{\em Nucl.~Phys.}\ {\bf B#1}}

\def \PhysR#1{{\em Phys.~Rev.}\ {\bf D#1}}

\def \PhysL#1{{\em Phys.~Lett.}\ {\bf #1B}}

\def \pseudo{pseu\-do\discretionary{-}{}{-}su\-per\-sym\-me\-try}

\renewcommand{\thefootnote}{\fnsymbol{footnote}}

\begin{document}

\begin{center}

{\LARGE \bf Effective Lagrangians\\[5mm] in Pseudo-Supersymmetry
   \footnote{Talk given at SUSY'02, HERA, Hamburg, June 2002.}}
\vskip .5in

{\bf Matthias Klein\footnote{E-mail: mklein@slac.stanford.edu.
      Research supported by the Deutsche Forschungsgemeinschaft.}}
\vskip 0.5cm
{\em SLAC, Stanford University, Stanford, CA 94309, USA.}

\end{center}

\vskip 1cm

\begin{center} {\bf ABSTRACT } \end{center}
I discuss effective field theories of brane-world models where different
sectors break different halves of the extended bulk supersymmetry.
It is shown how to consistently couple $\cN=2$ supersymmetric matter to 
$\cN=1$ superfields that lack $\cN=2$ partners but transform in a 
non-linear representation of the $\cN=2$ algebra. 
In particular, I explain how to couple an $\cN=2$ vector 
to $\cN=1$ chiral fields such that the second supersymmetry is
non-linearly realized.
This method is then used to study systems where different sectors break
different halves of supersymmetry, which appear naturally in models
of intersecting branes.

\setcounter{page}{1} \pagestyle{plain}
\renewcommand{\thefootnote}{\arabic{footnote}}
\setcounter{footnote}{0}


\section{Introduction}
D-brane models provide an interesting supersymmetry breaking mechanism
that I would like to study in an effective field theory approach 
\cite{pseudo}. Parallel
D-branes of the same dimensionality break half of the supersymmetry that
is present in the bulk. In the effective field theory on the D-brane 
world-volume, this is seen from the fact that the fields only fill
multiplets of the smaller supersymmetry algebra. Part of the supersymmetry 
is explicitly broken on the D-branes since the world-volume fields lack 
the corresponding superpartners. Supersymmetry can be completely broken by 
adding anti-D-branes (e.g., \cite{ADS,AU,AIQ,AADDS}), which preserve the 
other half of the bulk supersymmetry.
Thus, supersymmetry is broken in a non-local way in such models. Each sector 
taken separately preserves part of the supersymmetry.
A very similar situation arises in models containing stacks of D-branes
at angles that intersect each other (e.g., \cite{BGKL,quasi}). There is
an extended supersymmetry on each stack of D-branes, but only a fraction
of this supersymmetry is preserved at each intersection. Supersymmetry is
completely broken in models where different intersections break different
fractions of supersymmetry.

To determine the couplings of the bulk fields to the boundary fields, it
is important to note that the part of supersymmetry that is broken on the
D-branes is still non-linearly realized. Although a rigorous proof
for this statement is still missing, there is much evidence in favor of it,
including the following facts: (i) The $\cN=4$ vector multiplet of a
single D3-brane in flat ten-dimensional space contains just the right 
number of fermions to provide the goldstinos required for the non-linear
realization of the bulk $\cN=8$ supersymmetry. The counting still works
if the bulk supersymmetry is reduced to $\cN=4$ or $\cN=2$ by an orbifold
projection. (ii) Consider an $\cN=1$ supersymmetric $U(1)$ theory. The
requirement that the gaugino be the goldstino for a second non-linearly
realized supersymmetry uniquely determines the full non-linear action.
It turns out \cite{BG} that this action agrees with the supersymmetric
generalization of the Born-Infeld action, which is known to describe the
world-volume theory of D-branes. (iii) Consistent gravitino couplings
are very constrained and it is hard to imagine how the bulk gravitinos
corresponding to the broken supersymmetries can satisfy these constraints
without non-linear supersymmetry \cite{DM}.

In this talk, I want to discuss a four-dimensional toy model, where an
$\cN=2$ $U(1)$ bulk vector $(A_m,\lambda^{(1)},\lambda^{(2)},\phi)$
couples to two boundary sectors that preserve different halves of the 
$\cN=2$ bulk supersymmetry \cite{pseudo}. 
On the first boundary, there is an $\cN=1$ chiral multiplet 
$(\phi^{(1)},\psi^{(1)})$ which carries charge $q_1$ under the bulk vector
and an $\cN=1$ goldstino multiplet $(\lambda_g,\tilde A_m)$.
On the second boundary, there is an $\cN=1'$ chiral multiplet 
$(\phi^{(2)},\psi^{(2)})$ which carries charge $q_2$ under the bulk vector
and an $\cN=1'$ goldstino multiplet $(\lambda_g',\tilde A_m')$.
The boundary multiplets transform linearly under one supersymmetry and
non-linearly under the other supersymmetry. The bulk vector multiplet
transforms linearly under both supersymmetries.
\bea
\delta\phi^{(1)} &= &\sqrt2\,\xi^{(1)}\psi^{(1)}
   -2i\kappa\,\lambda_g\sigma^m
     \bar\xi^{(2)}\partial_m\phi^{(1)} \nonumber\\
\delta A_m &= &-i\lambda^{(1)}\sigma_m\bar\xi^{(1)}
   -i\lambda^{(2)}\sigma_m\bar\xi^{(2)} 
   \ +\  {\rm h.c.}\\
\delta\phi^{(2)} &= &\sqrt2\,\xi^{(2)}\psi^{(2)}
   -2i\kappa\,\lambda_g'\sigma^m
    \bar\xi^{(1)}\partial_m\phi^{(2)} \nonumber
\eea

To determine the Lagrangian describing this toy model, I will first
develop a method to couple $\cN=1$ multiplets to non-supersymmetric
matter and then generalize it to couplings of $\cN=2$ multiplets to
$\cN=1$ matter. Using the results of \cite{BG} on partially broken
$\cN=2$ supersymmetry, the \pseudo\ Lagrangian is then
easily obtained. A supergravity approach to \pseudo\ is pursued in
\cite{BFKQ}.

\section{Non-linear supersymmetry}
Let us start by briefly reviewing the formalism of non-linearly realized
supersymmetry. A supersymmetry transformation acts as a shift on the 
superspace coordinates:
\bea \label{coord_transf}
x^m &\to &x^{\prime m}
  =x^m-i(\xi\sigma^m\bar\theta-\theta\sigma^m\bar\xi) \nonumber\\
\theta &\to &\theta'=\theta+\xi \\
\bar\theta &\to &\bar\theta'=\bar\theta+\bar\xi \nonumber
\eea
An important observation is that the goldstino $\lambda_g$ of broken 
supersymmetry can be viewed as a hypersurface in superspace defined through
\be  \label{hypersurf}
\theta\ =\ -\kappa\,\lambda_g(x),
\ee
where $\kappa$ is a constant of mass dimension $-2$ related to the
supersymmetry breaking scale. This is very similar to the Goldstone
bosons of a spontaneously broken global internal symmetry being
hypersurfaces in the parameter space of the global symmetry group.

The requirement that the hypersurface (\ref{hypersurf}) be invariant
under supersymmetry transformations, $\theta'(x)=\theta(x')$,
implies the standard non-linear transformation law \cite{AV}
\be \label{gstino_transf}
\delta_\xi\lambda_g\ =\ {1\over\kappa}\xi-i\kappa\,
  (\lambda_g\sigma^m\bar\xi-\xi\sigma^m\bar\lambda_g)
  \partial_m\lambda_g
\ee
for the goldstino. This forms a non-linear realization of
the supersymmetry algebra \cite{AV}
\be \label{susy_alg}
[\delta_\eta,\delta_\xi]\ =\
       -2i\,(\eta\sigma^m\bar\xi-\xi\sigma^m\bar\eta)\,\partial_m.
\ee

A matter field $f(x)$ is well-defined on the hypersurface (\ref{hypersurf}) 
if, under a supersymmetry transformation, one has $f'(x)=f(x')$.
This implies
\be  \label{matter_transf}
\delta_\xi f\ =\ -i\kappa\,(\lambda_g\sigma^m\bar\xi-\xi\sigma^m\bar\lambda_g)
             \partial_m f,
\ee
which is the standard non-linear transformation for a matter field
in the goldstino background.

To construct an invariant action, one introduces covariant derivatives
\cite{IK}
\bea  \label{cov_deriv}
D_m &= & \left(\omega^{-1}\right)_m^{\ n}\partial_n, \\
\omega_m^{\ n} &=& \delta_m^{\ n}-i\kappa^2(
             \partial_m\lambda_g\sigma^n\bar\lambda_g
             -\lambda_g\sigma^n\partial_m\bar\lambda_g), \nonumber
\eea
and notes that $\det(\omega)$ transforms as a density under the
non-linear supersymmetry \cite{AV}. It is then straightforward to verify that
\be  \label{matter_action}
S=\int d^4x\,\det(\omega)\left(-{1\over2\kappa^2}-D_m f D^m f
           -V(f)\right)
\ee
is invariant under (\ref{gstino_transf}), (\ref{matter_transf}).

\section{Coupling $\cN=1$ to $\cN=0$}
Consider an $\cN=1$ superfield
\be  \label{Phi_expans}
\Phi(x,\theta,\bar\theta)\ =\ \phi(x)+\theta\psi(x)+\ldots
\ee
on which supersymmetry is linearly realized. To be able to couple the
lowest component $\phi$ of this superfield to a sector where supersymmetry
is non-linearly realized, we need to find a composite field $\hat\phi$
that transforms according to (\ref{matter_transf}) and reduces to $\phi$
in the limit $\kappa\to0$. It is easy to see that the desired composite
field is given by \cite{IK}
\be \label{phihat_def}
\hat\phi(x)\equiv\Phi(x,-\kappa\lambda_g(x),-\kappa\bar\lambda_g(x)).
\ee
Indeed, using the invariance of the hypersurface (\ref{hypersurf})
under linear supersymmetry transformations, one finds
\be \label{phihat_transf}
\delta_\xi\hat\phi\ =\ \hat\phi(x+\delta x)-\hat\phi(x)
   \ =\ -i\kappa(\lambda_g\sigma^m\bar\xi-\xi\sigma^m\bar\lambda_g)
         \partial_m\hat\phi.
\ee

As an illustrative example, consider a chiral multiplet $\Phi$. One has 
\be  \label{phihat} 
\hat\phi=\phi-\kappa\,\lambda_g\psi +\cO(\kappa^2).
\ee
The supersymmetric completion of the dilaton-like coupling
$(\phi^\dagger+\phi)F_{mn}F^{mn}$ is thus given by
\be  \label{dilcoupling}
\int d^4x
\left((\phi+\phi^\dagger-\kappa\,\lambda_g\psi-\kappa\,\bar\lambda_g\bar\psi)
        F_{mn}F^{mn}+\cO(\kappa^2)\right).
\ee

\section{Partially broken $\cN=2$}
Let us see how the above formalism of non-linearly realized
supersymmetry generalizes to partially broken $\cN=2$ supersymmetry.
The goldstino is now the lowest component of a superfield 
$\Lambda_g(x,\theta,\bar\theta)$ with respect to the unbroken supersymmetry.
I concentrate on the case, where the superpartner of the goldstino is a
$U(1)$ gauge boson:
\be \label{Lambda_expans}
\Lambda_{g\,\alpha}\ =\ -{i\over2}\left(\lambda_{g\,\alpha}+
               (\sigma^{mn}\theta)_\alpha F_{mn}+\ldots\right)
       \ =\ \hf W_\alpha+\cO(\kappa^2).
\ee
Under the second supersymmetry, $\Lambda_g$ transforms as 
\be  \label{sgstino_transf}
\delta^{(2)}\Lambda_g={1\over\kappa}\xi^{(2)}-i\kappa\,
            (\Lambda_g\sigma^m\bar\xi^{(2)}-\xi^{(2)}\sigma^m\bar\Lambda_g)
            \partial_m\Lambda_g.
\ee

The full non-linear invariant action for the goldstino superfield has been 
determined by the authors of \cite{BG}. They find
\be \label{sgstino_action}
S\ =\ \int d^4x\left[\qt\int d^2\theta\, W^2+
                     \qt\int d^2\bar\theta\, \bar W^2
    + {\kappa^2\over8}\int d^2\theta d^2\bar\theta\, W^2\bar W^2
    +\cO(\kappa^4)\right].
\ee

The bosonic terms of (\ref{sgstino_action}) coincide with the
Born-Infeld action,
\be \label{BI_action}
S_{\rm bos}\ =\ {1\over\kappa^2}\int d^4x\left(1-\sqrt{-\det(\eta_{mn}
               +\kappa\,F_{mn})}\right).
\ee

There is a chiral version of the non-linear supersymmetry transformation
which acts on chiral superfields $\tilde\Phi$ as
\be  \label{chir_transf}
\delta^{(2)}\tilde\Phi=-2i\kappa\,\tilde\Lambda_g\sigma^m\bar\xi^{(2)}
     \partial_m\tilde\Phi,
\ee
where $\tilde\Lambda_g(x,\theta,\bar\theta)=\Lambda(x^m-i\kappa^2\,
                       \Lambda_g\sigma^m\bar\Lambda_g,\theta,\bar\theta)$.

Defining $\Phi(x,\theta,\bar\theta)=\tilde\Phi(x^m+i\kappa^2\,\Lambda_g
           \sigma^m\bar\Lambda_g,\theta,\bar\theta)$, one finds that
an invariant action is of the form
\be  \label{phi_action}
S\ =\ 
   \int d^4x \left[\int d^2\theta d^2\bar\theta\,\hat E\,\Phi^\dagger\Phi
        +\int d^2\theta\,E_L\,\cP(\tilde\Phi)
        +\int d^2\bar\theta\,E_R\,\cP(\tilde\Phi^\dagger)\right],\nonumber
\ee
where
\bea
 \hat E &= &1+{\kappa^2\over8}\bar D^2\bar\Lambda_g^2
            +{\kappa^2\over8}D^2\Lambda_g^2 +\cO(\kappa^4), \nonumber\\
    E_L &= &1+{\kappa^2\over4}\bar D^2\bar\Lambda_g^2 
            +\cO(\kappa^4), \qquad E_R\ =\ E_L^\dagger.
\eea

\section{Coupling $\cN=2$ to $\cN=1$}
The method to couple an $\cN=2$ multiplet to $\cN=1$ matter is a 
straightforward generalization of the method explained in section 3.
In this talk, I will just give the result for the coupling of an
$\cN=2$ vector $(A_m,\lambda^{(1)},\lambda^{(2)},\phi)$ to $\cN=1$
matter. The $\cN=2$ vector can be split into an $\cN=1$ vector 
$V=(A_m,\lambda^{(1)})$ and an $\cN=1$ chiral multiplet
$\Phi=(\phi,\lambda^{(2)})$.

Using
\bea  \label{vec_transf}
\delta^{(2)} V &= &-{i\over\sqrt2}\,\theta\sigma^m\bar\theta
                      (\xi^{(2)}\sigma_m\bar D
                      \bar\Phi^\dagger + \bar\xi^{(2)}\bar\sigma^m D \Phi),
                  \nonumber\\
\delta^{(2)} \Phi &= &-i\sqrt2\,W\xi^{(2)},\\
\delta^{(2)} W_\alpha &= &{i\over\sqrt2}\,\bar\xi^{(2)}\bar DD_\alpha\Phi
                     -{i\over2\sqrt2}\xi^{(2)}_\alpha\bar D^2\Phi^\dagger,
                   \nonumber
\eea
one finds that
\be \label{hatphi_def}
                    \hat\Phi\equiv\Phi+i\sqrt2\,\kappa\,\tilde\Lambda_g W
                    -\qt\kappa^2\tilde\Lambda_g\tilde\Lambda_g\bar D^2
                     \Phi^\dagger
\ee
transforms as
$\delta^{(2)}\hat\Phi=-2i\kappa\,\tilde\Lambda_g\sigma^m\bar\xi^{(2)}
     \partial_m\hat\Phi$.
This implies that the $\cN=2$ supersymmetric generalization of the
dilaton-like coupling $\Phi\,\cW^\alpha\cW_\alpha$, where $\cW_\alpha$
is an $\cN=1$ Maxwell superfield without $\cN=2$ partner, is given by
\be  \label{phi_WW_action}
S\ =\ \int d^4x\,d^2\theta\,E_L\,\hat\Phi\,\cW^\alpha\cW_\alpha
                \ +\ h.c.
\ee

Similarly, one finds that
\be  \label{Vhat_def}
\hat V=V+i\kappa\,\theta\sigma^m\bar\theta\left(\bar\Lambda_g\bar\sigma_m D\Phi
          +\Lambda_g\sigma_m\bar D\Phi^\dagger\right)
        +\cO(\kappa^2).
\ee
transforms as
$\delta^{(2)}\hat V=-i\kappa\left(\Lambda_g\sigma^m\bar\xi^{(2)}
      -\xi^{(2)}\sigma^m\bar\Lambda_g\right) \partial_m\hat V$.
This implies that the $\cN=2$ supersymmetric generalization of the
gauge coupling term $\Phi_{\rm b}^\dagger\, e^V\,\Phi_{\rm b}$,
where $\Phi_{\rm b}$ is a chiral superfield without $\cN=2$ partner,
is given by
\be  \label{V_phi_action}
         S=\int d^4x d^2\theta d^2\bar\theta\,\hat E\,
                  \Phi_{\rm b}^\dagger\, e^{\hat V}\,\Phi_{\rm b}.
\ee

\section{Pseudo-Supersymmetry}
We are now in a position to write down the explicit Lagrangian for
the \pseudo\ toy model described in the introduction.
The field content of this model is summarized in table \ref{field_content}.
\begin{table}[th]\begin{center}
\begin{tabular}{|c|c|c|}
\hline
boundary 1  &bulk  &boundary 2 \\
$\cN=1$ matter  &$\cN=2$ vector  &$\cN=1'$ matter\\ \hline
$\tilde\Phi^{(1)}=(\phi^{(1)},\psi^{(1)})$ &$V=(A_m,\lambda^{(1)})$ &\\
             &$\Phi=(\phi,\lambda^{(2)})$ &\\
&{\rm or} &\\
&$V'=(A_m,\lambda^{(2)})$ &$\tilde\Phi^{(2)}=(\phi^{(2)},\psi^{(2)})$\\
&$\Phi'=(\phi,\lambda^{(1)})$ &\\
$\Lambda_g$ & &$\Lambda_g'$\\ \hline
\end{tabular}
\caption{\label{field_content}Field content of the \pseudo\ 
toy model. The $\cN=2$ vector can be split either into two $\cN=1$
multiplets or into two $\cN=1'$ multiplets.}
\end{center}\end{table}
Note that the $\cN=2$ vector can either be split into an $\cN=1$ vector $V$
and an $\cN=1$ chiral multiplet $\Phi$ or into an $\cN=1'$ vector $V'$
and an $\cN=1'$ chiral multiplet $\Phi'$. The chiral fields on the
first boundary couple to $V$ whereas the chiral fields on the second
boundary couple to $V'$. The Lagrangian is 
\be  \label{pseudo_Lagr}
\cL\ =\ \cL_{\rm bulk}\ +\ \cL^{(1)}\ +\ \cL^{(2)},  
\ee
with
\bea  \label{pseudo_LL}
\cL^{(1)} & =&\int d^2\theta d^2\bar\theta\,\hat E^{(1)}\,
               \Phi^{(1)\dagger}e^{q_1\,\hat V}\Phi^{(1)}
            \ +\ \int d^2\theta\, E_L^{(1)}\,
            \tilde\Lambda_g\tilde\Lambda_g\ +\ h.c.\nonumber\\
\cL_{\rm bulk} &= &\int d^2\theta d^2\bar\theta\,\Phi^\dagger e^V \Phi
                   \ +\ \qt\int d^2\theta\,WW + h.c.\\
\cL^{(2)} &= &\int d^2\tilde\theta d^2\bar{\tilde\theta}\,\hat E^{(2)}\,
               \Phi^{(2)\dagger}e^{q_2\,\hat V'}\Phi^{(2)}
                   \ +\ \int d^2\tilde\theta\, E_L^{(2)}\,
                   \tilde\Lambda_g'\tilde\Lambda_g'
                   \ +\ h.c.\nonumber
\eea

This supersymmetry breaking mechanism has several interesting consequences:
\begin{itemize}
\item There are no quadratic divergences at one-loop. Only Feynman diagrams
      involving fields from both boundaries can contribute to supersymmetry
      breaking. But such diagrams only arise at two-loop.
\item The scalar masses squared arising at two loops are expected to be 
      $\sim(g/4\pi)^4\,M^2$, where $g$ is the gauge coupling of the bulk 
      vector and $M$ is the cut-off scale of the effective field theory.
\item A non-vanishing vacuum energy only arises at three-loop.
      (See \cite{pseudoloop} for a calculation of scalar masses and
      vacuum energy in a 5D version of the model described in this talk.)
\item Both goldstinos stay massless (at tree-level) even when coupled to
      supergravity because the $U(1)$ gauge symmetries of their superpartners 
      are unbroken.
\item Gravitino masses arise only at three-loop.
\end{itemize}

\vskip5mm
\centerline{\bf Acknowledgements}

I would like to thank Michael Peskin for many stimulating discussions,  
Cliff Burgess, Elise Filotas and Fernando Quevedo for collaborating on
a related work which influenced my interest in pseudo-supersymmetry 
and Ignatios Antoniadis, Shamit Kachru and Albion Lawrence for answering
several questions I had.
Special thanks go to my wife for her support and encouragement.
This research is funded by the Deutsche Forschungsgemeinschaft.

\end{document}